\documentclass[twocolumn]{el-author}

\usepackage{epstopdf}

\newcommand{\ua}{\uparrow}
\newcommand{\nc}{\newcommand}
\makeatletter
\newcommand{\vast}{\bBigg@{1.77}}
\newcommand{\Vast}{\bBigg@{3.2}}
\newcommand{\vastl}{\bBigg@{4}}
\newcommand{\Vastl}{\bBigg@{5}}
\makeatother
\nc{\da}{\downarrow} \nc{\hc}{\hat{c}} \nc{\hS}{\hat{S}}
\nc{\bra}{\langle} \nc{\ket}{\rangle} \nc{\eq}{equation (\ref}
\nc{\h}{\hat} \nc{\hT}{\h{T}}\nc{\be}{\begin{eqnarray}}
\nc{\ee}{\end{eqnarray}}\nc{\rd}{\textrm{d}}\nc{\e}{eqnarray}\nc{\hR}{\hat{R}}\nc{\Tr}{\mathrm{Tr}}
\nc{\tS}{\tilde{S}}\nc{\tr}{\mathrm{tr}}\nc{\8}{\infty}\nc{\lgs}{\bra\ua,\phi|}\nc{\rgs}{|\ua,\phi\ket}
\nc{\hU}{\hat{U}}\nc{\lfs}{\bra\phi|}\nc{\rfs}{|\phi\ket}\nc{\hZ}{\hat{Z}}\nc{\hd}{\hat{d}}\nc{\mD}{\mathcal{D}}
\nc{\bd}{\bar{d}}\nc{\bc}{\bar{c}}\nc{\mc}{\mathcal}\nc{\ea}{eqnarray}\nc{\mG}{\mathcal{G}}\nc{\bce}{\begin{center}}
\nc{\ece}{\end{center}}
\date{16th March 2019}

\begin{document}

\title{Effective rate analysis over Fluctuating Beckmann fading channels}

\author{Hussien Al-Hmood and H.S. Al-Raweshidy}

\abstract{The effective rate of Fluctuating Beckmann (FB) fading channel is analysed. The moment generating function (MGF) of the instantaneous signal-to-noise (SNR) is used first to derive the effective rate for arbitrary values of the fading parameters in terms of the extended generalised bivariate Meijer's-$G$ function (EGBMGF). 
For integer valued of the multipath and shadowing severity fading parameters, the probability density function (PDF) of the instantaneous SNR is employed. To this end, simple exact mathematically tractable analytic expression is obtained. The Monte Carlo simulations and the numerical results are
presented to verify the validation of our analysis.}

\maketitle

\section{Introduction}
The effective rate (ER) has been proposed to measure the performance of the wireless communication systems under the quality of service (QoS) constraints, such as system delays, that have not been taken into consideration by Shannon [1]. Accordingly, the analysis of this performance metric over fading channels has been given a special attention by several works. For instance, in [2], the ER over Nakagami-$n$, namely, Rician, fading channel is analysed using the moment generating function (MGF) of the instantaneous signal-to-noise (SNR). 
\par Recently, several efforts have been achieved to study the ER over different generalised fading channels. This is because these channels include most of the classic fading models such as Nakagami-$m$ as special cases but with better fitting to the practical measurements. For example, the ER over $\kappa-\mu$ and $\eta-\mu$ fading channel which are used to model the line-of-sight (LoS) and non-LoS (NLoS) communication scenarios are investigated in [3] and [4], respectively. In [5], the expression of the ER over $\kappa-\mu$ shadowed fading condition which is composite of $\kappa-\mu$ fading and Nakagami-$m$ distributions is provided in terms of the extended generalised bivariate Meijer's-$G$ function (EGBMGF) which doesn't give clear insights about the results against the variation of the fading parameters. The analysis in [6] is carried out over composite $\alpha-\eta-\mu$/gamma fading channels using two approximate unified frameworks. However, the derived expressions are also included the EGBMGF. The Fisher-Snedecor $\mathcal{F}$ distribution is used to study the ER over composite multipath/shadowed fading condition in [7]. Although, the expression is given in terms of a single variable Meijer's-$G$ function, this fading model includes few number of the conventional distributions as special cases.  
\par More recent, the Fluctuating Beckmann (FB) fading channel has been proposed as an extension of the $\kappa-\mu$ shadowed and the classical Beckmann fading models [8]. In addition, it includes as special cases the one-sided Gaussian, Rayleigh, Nakagami-$m$, Rician, $\kappa-\mu$, $\eta-\mu$, $\eta-\kappa$, Beckmann, Rician shadowed and the $\kappa-\mu$ shadowed distributions. Accordingly, this letter is devoted to analyse the ER over FB fading channel. To the best of the authors' knowledge, there is no effort has been dedicated to investigate the aforementioned analysis in the open literature. To this end, novel exact mathematically tractable expression is derived in terms of the EGBMGF using the MGF approach. To gain more insights into the impact of the channel parameters on the ER, the PDF is utilised to provide novel simple exact closed-form expression via assuming the fading parameters are integer values. 
 
\section{Effective rate}
The normalised ER is evaluated  by [4, eq. (1)]
\label{eqn_1}
\begin{equation}
\mathcal{R}=-\frac{1}{A}\mathrm{log}_2\big(\mathbb{E}\{(1+\gamma)^{-A}\}\big)
\end{equation}
where $\mathbb{E}\{.\}$ denotes the expectation, $\gamma$ is the instantaneous signal-to-noise ratio (SNR), and $A \triangleq \theta TB/\mathrm{ln}2$ with $\theta$, $T$, and $B$ are the delay exponent, block duration, and bandwidth of the system, respectively.

\section{Fluctuating Beckmann fading channel}
The MGF of $\gamma$ over FB fading channel model is expressed as [8, eq. (3)]
\label{eqn_2}
\begin{align}
\mathcal{M}_{\gamma}(s)=\frac{1}{(\alpha_1 c_1 c_2)^m} \Big(1+\frac{\eta \bar{\gamma}}{\Omega}s \Big)^{m-\frac{\mu}{2}}  \Big(1+\frac{\bar{\gamma}}{\Omega}s \Big)^{m-\frac{\mu}{2}}  \nonumber\\ 
\Big(1+\frac{\bar{\gamma}}{c_1}s \Big)^{-m}  \Big(1+\frac{\bar{\gamma}}{c_2}s \Big)^{-m}
\end{align}
where $\Omega=\frac{\mu (1+\eta)(1+\kappa)}{2}$, $\mu$ is the real extension of the multipath clusters, $\bar{\gamma}$ is the average SNR, and $m$ is the shadowing severity index. Moreover, $\kappa=\frac{p^2+q^2}{\mu(\sigma^2_{x}+\sigma^2_{y})}$, $\eta = \frac{\sigma^2_{x}}{\sigma^2_{y}}$, $\sigma^2_{x}=\mathbb{E}[X^2_{i}]$, $\sigma^2_{y}=\mathbb{E}[Y^2_{i}]$, and $p_{i}$ and $q_{i}$ are real numbers for $i$th cluster. The parameters $c_1$ and $c_2$ are the roots of $\alpha_{1} s^2+ \beta s+1$ with [8, eqs. (7-8)]
\label{eqn_3}
\begin{align}
\alpha_{1}&=\frac{\eta}{\Omega^2}+\frac{\kappa  (\varrho^2+\eta)}{m  \Omega (1+\varrho^2)(1+\kappa)} \nonumber\\
\beta&=-\frac{1}{1+\kappa}\bigg[\frac{2}{\mu}+\frac{\kappa}{m}\bigg]
\end{align}
where $\varrho^2=\frac{p^2}{q^2}$.
\par When $\mu$ and $m$ are even and integer numbers, respectively, the PDF of $\gamma$ is given as [8, eq. (10)]
\label{eqn_4}
\begin{equation}
f_{\gamma}(\gamma)=\frac{1}{ \alpha^m_1 \bar{\gamma}^\mu} \Big(\frac{\eta}{\Omega^2}\Big)^{m-\frac{\mu}{2}}\sum_{i=1}^{N(m,\mu)} e^{-\frac{\vartheta_i}{\bar{\gamma}}\gamma} \sum_{j=1}^{|\omega_{i}|}\frac{A_{ij}}{(j-1)!}\gamma^{j-1}
\end{equation}
where $\omega=[m, m, \frac{\mu}{2}-m, \frac{\mu}{2}-m]$, $\vartheta=[c_1, c_2, \frac{\Omega}{\eta}, \Omega]$, $N(m,\mu)= 2[1+\text{u}(\frac{\mu}{2}, m)]$, u(.) is the unit step function, and $A_{ij}$ is computed by [7, eq.(51)].
\section{Effective rate using MGF approach}
According to [2, eq. (3)], (1) can be computed by the MGF of $\gamma$ as follows 
\label{eqn_5}
\begin{equation}
\mathcal{J}=\mathbb{E}\{(1+\gamma)^{-A}\}=\frac{1}{\Gamma(A)} \int_0^\infty s^{A-1} e^{-s} \mathcal{M}_\gamma(s)ds
\end{equation}
where $\Gamma(a)=\int_0^\infty x^{a-1} e^{-x} dx$ is the Gamma function.
\par Substituting (1) in (5) to yield 
\label{eqn_6}
\begin{align}
\mathcal{J}&=\frac{1}{(\alpha_1 c_1 c_2)^m \Gamma(A)} \int_0^\infty s^{A-1} e^{-s} \Big(1+\frac{\eta \bar{\gamma}}{\Omega}s \Big)^{m-\frac{\mu}{2}}  \nonumber\\
&\times  
\Big(1+\frac{\bar{\gamma}}{\Omega}s \Big)^{m-\frac{\mu}{2}} 
\Big(1+\frac{\bar{\gamma}}{c_1}s \Big)^{-m}  \Big(1+\frac{\bar{\gamma}}{c_2}s \Big)^{-m} d{s}
\end{align}
\par The following identity [9, eq. (10)] can be used in (6)
\label{eqn_5}
\begin{equation}
(1+x)^a =\frac{1}{\Gamma(-a)}
G^{1,1}_{1,1} \vast[\begin{matrix}
    1+a\\
  0\\ 
\end{matrix}\vast\vert x
\vast]
\end{equation}
\par Accordingly, we have
\label{eqn_7}
\begin{align}
\mathcal{J}=\frac{1}{(\alpha_1 c_1 c_2)^m \Gamma(A) [\Gamma(m) \Gamma(\frac{\mu}{2}-m)]^2}  \hspace{3 cm} \nonumber\\
\times \int_0^\infty s^{A-1} e^{-s}  
G^{1,1}_{1,1} \vast[\begin{matrix}
   1+m-\frac{\mu}{2}\\
  0\\ 
\end{matrix}\vast\vert \frac{\eta \bar{\gamma}}{\Omega}s
\vast]
G^{1,1}_{1,1} \vast[\begin{matrix}
   1+m-\frac{\mu}{2}\\
  0\\ 
\end{matrix}\vast\vert \frac{\bar{\gamma}}{\Omega}s
\vast]  \nonumber\\ 
G^{1,1}_{1,1} \vast[\begin{matrix}
   1-m\\
  0\\ 
\end{matrix}\vast\vert \frac{\bar{\gamma}}{c_1}s
\vast]
G^{1,1}_{1,1} \vast[\begin{matrix}
   1-m\\
  0\\ 
\end{matrix}\vast\vert \frac{\bar{\gamma}}{c_2}s
\vast]
 d{s}
\end{align}
\par Using the integral representation of the Meijer's-G function [9, eq. (5)] in (8) to yield
\label{eqn_9}
\begin{align}
\mathcal{J}=\frac{1}{(\alpha_1 c_1 c_2)^m \Gamma(A) [\Gamma(m) \Gamma(\frac{\mu}{2}-m)]^2} \nonumber \hspace{2.5 cm}\\
\times \frac{1}{(2 \pi j)^4} \int_{\mathcal{R}_1} \cdots \int_{\mathcal{R}_4} \Big(\prod_{i=1}^4 \Gamma(r_i)\Big) \Gamma\Big(\frac{\mu}{2}-m-r_1 \Big) \Gamma\Big(\frac{\mu}{2}-m-r_2\Big) \nonumber\hspace{-0.5 cm}\\
\Gamma(m-r_3) \Gamma(m-r_4)\int_0^\infty s^{A-\sum_{i=1}^4 r_i-1} e^{-s} ds dr_1 dr_2 dr_3 dr_4
\end{align}
where $j=\sqrt{-1}$ and $\mathcal{R}_i$ for $i=1,\cdots,4$ is the $i^\text{th}$ suitable closed contours in the complex $r$-plane.
\par With the help of [10, eq. (1.1.6)], the inner integral of (9)
\label{eqn_10}
\begin{align}
\mathcal{J}=\frac{1}{(\alpha_1 c_1 c_2)^m \Gamma(A) [\Gamma(m) \Gamma(\frac{\mu}{2}-m)]^2} \nonumber \hspace{2.7 cm}\\
\times \frac{1}{(2 \pi j)^4} \int_{\mathcal{R}_1} \cdots \int_{\mathcal{R}_4} \Big(\prod_{i=1}^4 \Gamma(r_i)\Big) \Gamma\Big(\frac{\mu}{2}-m-r_1 \Big) \Gamma\Big(\frac{\mu}{2}-m-r_2\Big) \nonumber\hspace{0.2 cm}\\
\Gamma(m-r_3) \Gamma(m-r_4) \Gamma(A-r_1-r_2-r_3-r_4) dr_1 dr_2 dr_3 dr_4
\end{align}
\par It can be observed that (10) can be expressed in exact closed-form in terms of the EGBMGF as follows
 \label{eqn_11}
\begin{align}
\mathcal{J}=\frac{1}{(\alpha_1 c_1 c_2)^m \Gamma(A) [\Gamma(m) \Gamma(\frac{\mu}{2}-m)]^2} \nonumber \hspace{2.7 cm}\\
\times G^{0,1:1,1;1,1;1,1;1,1}_{1,0:1,1;1,1;1,1;1,1} \Big[\begin{matrix}
    1-A\\
  -\\ 
\end{matrix}\Big\vert
\begin{matrix}
    1+m\frac{\mu}{2}\\
  0\\ 
\end{matrix}\Big\vert  \nonumber \hspace{3 cm}\\
\begin{matrix}
   1+m-\frac{\mu}{2}\\
  0\\ 
\end{matrix}\Big\vert
\begin{matrix}
   1-m\\
  0\\ 
\end{matrix}\Big\vert
\begin{matrix}
   1-m\\
  0\\ 
\end{matrix}\Big\vert
\frac{\eta \bar{\gamma}}{\Omega}, \frac{\bar{\gamma}}{\Omega}, \frac{\bar{\gamma}}{c_1},\frac{\bar{\gamma}}{c_2}
\Big]
\end{align}
\par One can see that the EGBMGF is not available in MATLAB and MATHEMATICA software packages. Therefore, this function has been calculated in this letter by employing a MATHEMATICA code that is implemented in [12, Table II].
\section{Effective rate using PDF approach} The expectation of (1) can be evaluated by the PDF as follows 
\label{eqn_12}
\begin{equation}
\mathcal{J}=\int_0^{\infty} (1+\gamma)^{-A} f_\gamma(\gamma) d\gamma
\end{equation}
\par When $m$ and $\mu$ are integer and even numbers, respectively, $\mathcal{J}$ of (12) can be computed by plugging (4) in (12). Thus, this yields 
\label{eqn_13}
\begin{align}
\mathcal{J}=\frac{1}{ \alpha^m_1 \bar{\gamma}^\mu} \Big(\frac{\eta}{\Omega^2}\Big)^{m-\frac{\mu}{2}} \sum_{i=1}^{N(m,\mu)}  \sum_{j=1}^{|\omega_{i}|}\frac{A_{ij}}{(j-1)!} \nonumber\\
\times \int_0^{\infty} \gamma^{j-1} (1+\gamma)^{-A} e^{-\frac{\vartheta_i}{\bar{\gamma}}\gamma} d\gamma
\end{align}
\par With the aid of [10, eq. (1.3.14), pp. 38], the integral in (13) can be calculated in exact closed-form expression as follows
\label{eqn_14}
\begin{equation}
\mathcal{J}=\frac{1}{ \alpha^m_1 \bar{\gamma}^\mu} \Big(\frac{\eta}{\Omega^2}\Big)^{m-\frac{\mu}{2}}  \sum_{i=1}^{N(m,\mu)}  \sum_{j=1}^{|\omega_{i}|} A_{ij} U\Big(j;j-A+1;\frac{\vartheta_i}{\bar{\gamma}} \Big)
\end{equation}
where $U(.)$ is the Tricomi hypergeometric function of the second kind defined in [10, eq. (1.3.15), pp. 38].
\begin{figure}[h]
\centering
  \includegraphics[width=3.45 in, height=2.5 in]{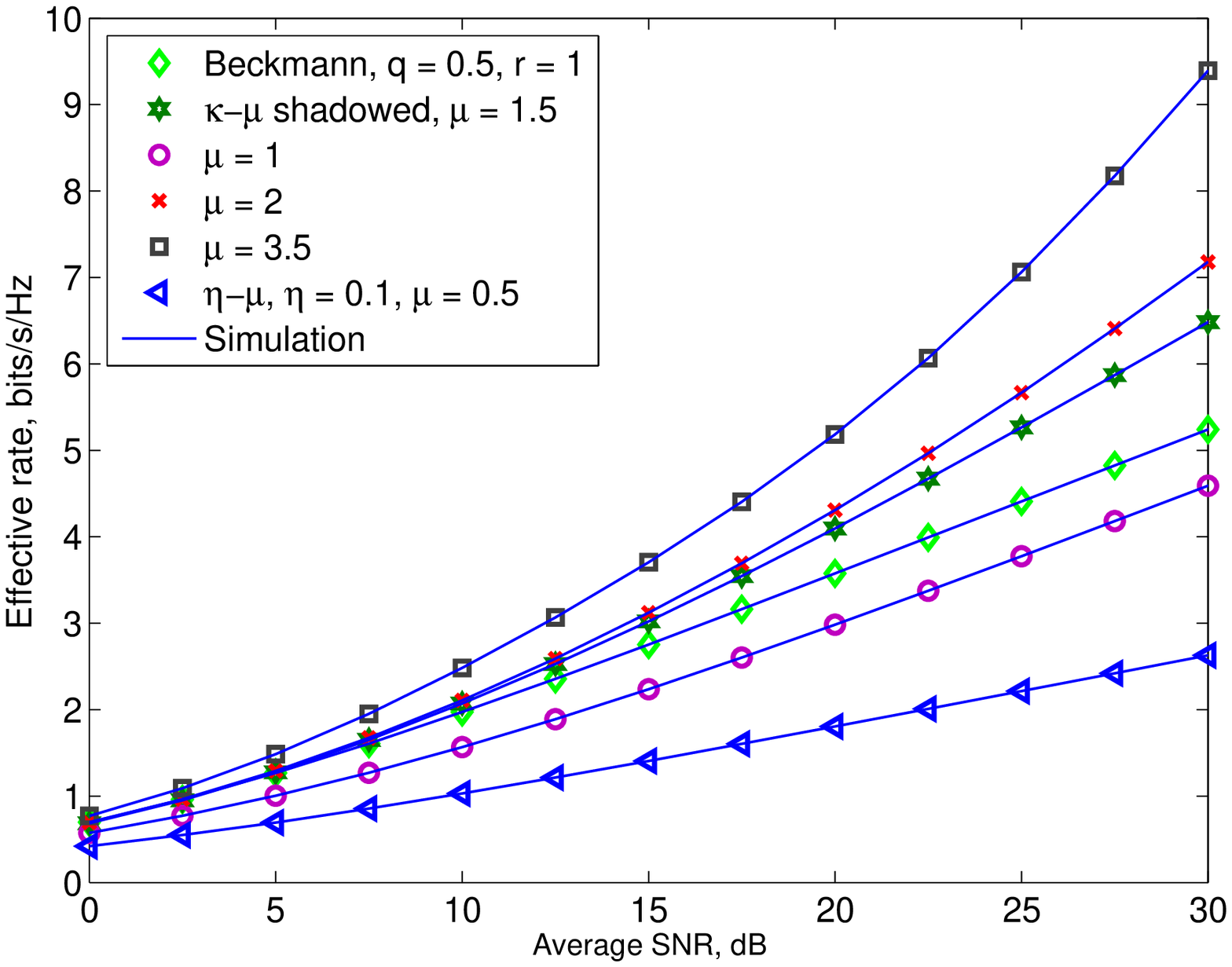} 
\centering
\caption{Normalised ER against the average SNR for $m = 1$, $K = 1$,  $\eta = 0.1$, $\varrho^2 = 0.1$, $A = 2$ and different values of $\mu$.}
\end{figure} 
\begin{figure}[h]
\centering
\includegraphics[width=3.45 in, height=2.5 in]{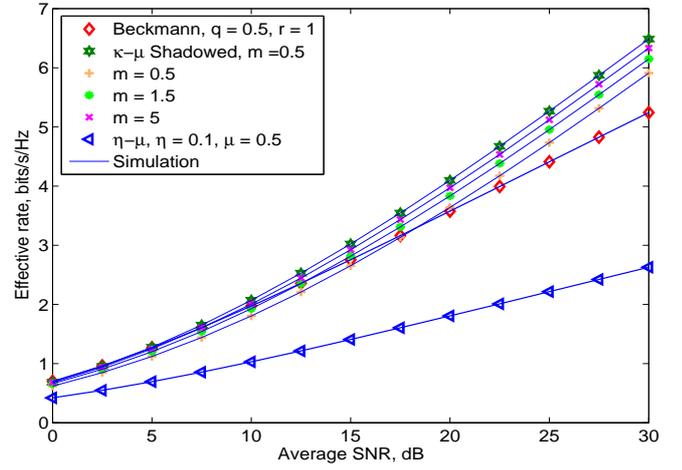} 
\centering
\caption{Normalised ER against the average SNR for $\mu = 1.5$, $K = 1$,  $\eta = 0.1$, $\varrho^2 = 0.1$, $A = 2$ and different values of $m$.}
\end{figure}

\section{Numerical results} 
The numerical and simulation results for the ER against average SNR for $K = 1$,  $\eta = 0.1$, $\varrho^2 = 0.1$, $A = 2$ and different values of $\mu$ and $m$ are presented in Fig. 1 and Fig. 2, respectively. In both figures, when $\mu$ or/and $m$ increase, the performance of the ER becomes better. This refers to a large number of the multipath clusters and less shadowing effect at the receiver, respectively. In addition, the ER over Beckmann fading channel that has not been done in the open literature is also explained via inserting $m \rightarrow\infty$, $\mu = 1$, and $r = \varrho^2 = 1$ in (11) or (14). Additionally, the provided results demonstrate the performance of the ER over $\eta-\mu$ and $\kappa-\mu$ shadowed fading channels that are deduced via using specific values for the fading parameters of FB fading model [8, Table I]. 

\section{Conclusion}
The ER over FB fading channel model which includes wide range of the fading distributions has been analysed using two different exact expressions. The MGF approach is employed first to derive the ER for arbitrary values of $m$ and $\mu$ where its expressed in terms of the EGBMGF. In the second case, $m$ and $\mu$ are assumed to be integer and even numbers, respectively and the PDF is used to obtain simple exact closed-form analytic expression of the ER. The results are provided for different scenarios via utilising various values of the fading parameters as well as the special cases of the FB fading channel.   

\vskip3pt
\copyright
\vskip5pt

\noindent Hussien Al-Hmood (\textit{Electrical and Electronics Engineering Department, Thi-Qar University, Thi-Qar, Iraq})
\vskip3pt

\noindent E-mail: hussien.al-hmood@brunel.ac.uk

\noindent H.S. Al-Raweshidy (Electronic and Computer Engineering
Department, College of Engineering, Design and Physical Sciences, Brunel University, London, United Kingdom)

\end{document}